# Relativistic deflection of background starlight measures the mass of a nearby white dwarf star

Kailash C. Sahu[1]*, Jay Anderson[1], Stefano Casertano[1], Howard E. Bond[2], Pierre Bergeron[3,] Edmund P. Nelan[1], Laurent Pueyo[1], Thomas M. Brown[1], Andrea Bellini[1], Zoltan G. Levay[1], Joshua Sokol[1], Martin Dominik[4], Annalisa Calamida[1], Noé Kains[1], Mario Livio[5]

**Affiliations:**

[1]Space Telescope Science Institute, 3700 San Martin Drive, Baltimore, MD 21218, USA.
[2]Department of Astronomy and Astrophysics, Pennsylvania State University, University Park, PA 16802, USA.
[3]Département de Physique, Université de Montréal, C.P. 6128, Succ. Centre-Ville, Montréal, QC H3C 3J7, Canada.
[4]University of St Andrews, School of Physics and Astronomy, North Haugh, St. Andrews, KY16 9SS, UK.
[5]Department of Physics and Astronomy, University of Nevada, Las Vegas, 4505 South Maryland Parkway, Las Vegas, NV 89154, USA
*Corresponding author. Email: ksahu@stsci.edu.

**Abstract**: Gravitational deflection of starlight around the Sun during the 1919 total solar eclipse provided measurements that confirmed Einstein's general theory of relativity. We have used the *Hubble Space Telescope* to measure the analogous process of astrometric microlensing caused by a nearby star, the white dwarf Stein 2051 B. As Stein 2051 B passed closely in front of a background star, the background star's position was deflected. Measurement of this deflection at multiple epochs allowed us to determine the mass of Stein 2051 B —the sixth nearest white dwarf to the Sun—as 0.675±0.051 solar masses. This mass determination provides confirmation of the physics of degenerate matter and lends support to white dwarf evolutionary theory.

**One Sentence Summary:** A century after Einstein postulated general relativity, bending of light by a white dwarf is used to determine its mass.

**O**ne of the key predictions of general relativity set forth by Einstein (*1*) was that the curvature of space near a massive body causes a ray of light passing near it to be deflected by twice the amount expected from classical Newtonian gravity. The subsequent experimental verification of this effect during the 1919 total solar eclipse (*2,3*) confirmed Einstein's theory, which was declared "one of the greatest—perhaps the greatest—of achievements in the history of human thought" (*4*).

In a paper in this journal 80 years ago, Einstein (*5*) extended the concept to show that the curvature of space near massive objects allows them to act like lenses, with the possibility of substantially increasing the apparent brightness of a background star. Despite Einstein's pessimistic view that "there is no hope of observing this phenomenon directly" *(5),* the prospect of detecting dark matter through this effect (*6*), now known as microlensing, revived interest in this subject. Coupled with improvements in instrumentation, this led to the detection of large numbers of microlensing brightening events in the Galactic bulge (*7*), the Magellanic Clouds (*8,9*), and the Andromeda Galaxy (*10*). Monitoring of these events has led to the discovery of several extrasolar planets (*11,12*). Other forms of gravitational lensing by intervening massive galaxies and dark matter produce multiple or distorted images of background galaxies *(13).*

Within the Milky Way, all microlensing encounters discovered so far have been brightening events. No shift in the apparent position of a background star caused by an intervening massive body has been observed outside the solar system—which is not surprising, because the deflections are tiny. Even for the nearest stars, the angular offset is two to three orders of magnitude smaller than the deflection of 1.75 arcsec measured during the 1919 solar eclipse.

**Relativistic deflections by foreground stars**

When a foreground star (the lens) is perfectly superposed on a background star (the source), the lensed image of the source will form a circle, called the Einstein ring. The angular radius of the Einstein ring is *(14)*

$$\theta_E = \sqrt{4GM \ / \ c^2 D_r} \ , \tag{1}$$

where $M$ is the lens mass and $D_r$ is the reduced distance to the lens, given by $1/D_r = 1/D_l - 1/D_s$, $D_l$ and $D_s$ being the distances to the lens and to the source, respectively. For typical cases like the Galactic bulge and Magellanic Clouds brightening events, the radius of the Einstein ring is less than a milliarcsecond (mas). However, for very nearby stellar lenses, it can be as large as tens of mas.

In the more general case where the lens is not exactly aligned with the source, the source is split into two images, the minor image lying inside and the major image outside the Einstein ring. The major image is always the brighter, with the brightness contrast increasing rapidly as the lens-source separation increases. In practical cases of lensing by stars, the two images either cannot be resolved, or the minor image is too faint to be detected. In both cases, the net effect is an apparent shift in the centroid position of the source. This phenomenon is referred to as astrometric microlensing (*15*).

In cases where the angular separation between the lens and the source is large compared to $\theta_E$, so that the minor image is well resolved but is too faint and too close to the bright lens to be detected, only the major image can be monitored. In that situation, the change in angular position of the source caused by the deflection of the light rays, $\delta\theta$, can be expressed as *(16)*

$$\delta\theta = 0.5 \left[ \sqrt{(u^2 + 4)} - u \right] \theta_E, \tag{2}$$

where $u = \Delta\theta/\theta_E$, and $\Delta\theta$ is the lens-source angular separation. Equations 1 and 2 show that the mass of the lens can be determined by measuring the deflection of the background source's position at a known angular separation from the lens, provided the reduced distance to the lens is known. Astrometric microlensing thus provides a technique for direct determination of stellar masses, in those favorable cases of a nearby star fortuitously passing close in front of a distant background source. Unlike classical methods involving binaries, this method can be applied to mass measurements for single stars.

**Predicting astrometric microlensing events due to nearby stars**

We carried out a large-scale search for events in which nearby stars with large proper motions (PMs) would pass closely in front of background sources. We used an input PM catalog (17) of ~5,000 stars, with updated positions and PMs based on modern sky-survey data (18-19). Parallaxes were also included when available. We then projected the positions of all ~5,000 stars forward and searched for close passages near fainter background stars contained in the Guide Star Catalog version 2.3 (20). One of the predicted events was a close passage of the nearby white dwarf (WD) star Stein 2051 B in front of a background star with $V$-band magnitude of 18.3, located at right ascension (RA) = 4:31:15.004 and declination (Dec) = +58:58:13.70 (J2000 equinox). This encounter was also predicted in an independent list of upcoming microlensing events (21). We estimated that the closest encounter would occur during March 2014, with an impact parameter of $0''.1$.

**The Stein 2051 binary system**

Stein 2051 is a nearby visual binary whose brighter but less-massive component, Stein 2051 A, is an 11th-magnitude ($V$ band) main-sequence star of spectral type M4 (19). The 12.4-magnitude companion, Stein 2051 B (also known as WD 0426+588), is the sixth-nearest known WD (22). It is currently at an angular separation of ~$10''.1$ from Stein 2051 A.

Stein 2051 B is the nearest and brightest known featureless-spectrum WD of spectral type DC, having a helium-rich photosphere. We determined its effective temperature to be $T_{\text{eff}} = 7122 \pm 181$ K, based on calibrated broad-band photometry and model atmospheres (23,29). Combining the photometry and temperature with the measured parallax (discussed below), we find the radius of the WD to be $0.0114 \pm 0.0004$ solar radii (R$_\odot$). At this radius, Stein 2051 B would be expected to have a mass of ~0.67 solar masses (M$_\odot$) if it obeys a normal mass-radius relation for carbon-oxygen (CO) core WDs (24).

For these parameters, the WD's Einstein ring would have a radius of about 31 mas. At a separation of $0''.5$ from Stein 2051 B, the background star would be displaced by ~2 mas. Actual measurement of such a deflection, especially so close to the glare of the bright foreground star, would be extremely challenging for seeing-limited ground-based telescopes. However, the measurement is within the capabilities of the instruments on the *Hubble Space Telescope* (HST).

The actual mass of Stein 2051 B has been a matter of debate. Photographic observations extending back to 1908 have been used to claim departures from linear PMs of A and B (25,26), implying a detection of orbital motion and a mass ratio of $M_B/M_A = 2.07$. Assuming A to be a normal M4 main-sequence star of 0.24 M$_\odot$, the mass of B was estimated (25,26) to be 0.50 M$_\odot$. A mass this low, combined with its inferred radius, would lead to the requirement for the WD to have an iron core. Such a result would be in conflict with normal single-star evolution, in which the stellar core only undergoes hydrogen and then helium fusion, resulting in a CO core composition for the WD remnant (*24*). Moreover, a WD cooling age of ~2.0 Gyr derived for Stein 2051 B (*22*), combined with the implied long main-sequence lifetime of the progenitor of a low-mass WD (*27*), would give the system a total age uncomfortably close to the age of the Universe. However, the detection of non-linear PMs was not confirmed by subsequent measurements *(28)*, implying that the orbital period of the A-B pair exceeds ~1000 yr, and appearing to invalidate the earlier mass determination.

### *Hubble Space Telescope* observations and analysis

We imaged the field of Stein 2051 with the Wide Field Camera 3 at eight epochs (denoted E1 through E8) between October 2013 and October 2015. The HST observing log is given in Table 1. We employed a range of exposure times (*29*), depending on how close the WD was to the source star, in three filters: F606W (a wide *V* band), F763M (a medium red band), and F814W (equivalent to *I* band). We used the long-exposure broad-band F606W and F814W images for the deflection measurements of the source star. For determining the location of the WD we used short-exposure F606W and F814W images and all the F763M frames (in which the WD did not saturate the detector).

Figure 1 shows a color image of the region around Stein 2051 B, created by superposing F606W and F814W frames at epoch E1. The path of the WD past the source, due to PM and parallax, is depicted by the wavy line. Closest approach to the source star occurred on 5 March 2014, at an angular separation of 103 mas. Even at closest separation, the photometric microlensing amplification would be only 1.0% and thus swamped by light from the bright WD, so we did not attempt to measure it. For measurements of the deflection, we obtained data at separations ranging from 203 to 3897 mas. (The amplification at our observed minimum separation of 203 mas is even less, 0.1%, which is undetectable.) We used the observations at all eight epochs to determine the parallax and PM of the WD. However, the presence of the 400 times brighter WD adjacent to the faint source made the deflection measurements possible only at separations larger than ~450 mas. In the epoch E2 frames, the source lay on a diffraction spike of the WD, and consequently the measurements had large uncertainties of ~0.1 pixels. We thus used only the observations taken at epochs E1, E6, E7, and E8 for the deflection analysis.

Full details of our data-analysis procedures are given in the online supplementary material (*29*), but we summarize them here. We used the flat-fielded images produced by the Space Telescope Science Institute pipeline reductions (*30*) for the analysis. We used empirical effective point spread functions (PSFs) (*31*) to measure the positions of the 26 reference stars in the surrounding field, and then applied distortion corrections (*32*) to convert the measured positions into locations in an undistorted reference frame. We used empirical PSFs to determine the source positions at epochs E1, E7 and E8, during which the source was sufficiently separated from the WD that it did not suffer from any contamination from the WD. At E6, when the source was 11.3 pixels away from the WD, the contamination from the WD was estimated to be about 5% of the source flux. We therefore performed an optimal PSF subtraction of the WD before measuring the position of the source (*29*).

We estimated distances to the reference stars and source by obtaining their magnitudes and colors determined from our HST frames, the Panoramic Survey Telescope and Rapid Response System (Pan-STARRS) survey (*33*), and the Two Micron All Sky Survey (2MASS) (*34*). Theoretical stellar models (*35*) were then used to determine individual distances of the reference stars, which lie in the range ~0.8 to 2.1 kiloparsecs (kpc). The source itself is estimated to be a K dwarf at a distance of ~2.0 kpc.

To establish a fixed reference frame, we first determined the PM of each reference star with respect to the ensemble, using an iterative procedure (*29*). Then at each epoch the reference-star positions were corrected for PM, and the positions of the source and the WD were determined relative to this adjusted frame. The estimated uncertainty in the position of the source star relative to the adjusted frame is ~0.4 mas in each individual exposure.

**Parallax and proper motion of Stein 2051 B**

From our measurements of Stein 2051 B at the eight epochs, we find a parallax of $180.7 \pm 1.0$ mas relative to the background frame, or an absolute parallax of $181.5 \pm 1.0$ mas after correction by the mean of our estimated reference-star parallaxes of $0.8 \pm 0.2$ mas. Absolute parallaxes for the Stein 2051 system have been measured previously (*36-38*), giving a weighted mean of $180.9 \pm 0.5$ mas, in statistical agreement with our result. The corresponding distance is $5.52 \pm 0.01$ parsecs (pc). We measure PM components for Stein 2051 B of $(\mu_\alpha, \mu_\delta) = (+1336.3 \pm 1.0, -1947.5 \pm 1.0)$ mas yr$^{-1}$, where $\mu_\alpha$ is the PM in the RA direction and $\mu_\delta$ is the PM in the Dec direction. These again are not absolute, but relative to our background reference frame. The PM of Stein 2051 B relative to a different selection of nearby reference stars has been measured previously to be $(\mu_\alpha, \mu_\delta) = (+1361.8 \pm 2.0, -1930.4 \pm 2.0)$ mas yr$^{-1}$ (*36*), and the absolute PM to be $(\mu_\alpha, \mu_\delta) = (+1335.6 \pm 2.5, -1962.6 \pm 2.5)$ mas yr$^{-1}$ (*39*). The differences between these values show sensitivity to the bulk motion of the chosen reference frame, but do not affect our interpretation of the event as long as our measurements are consistently in the same reference frame for both the source and the WD (*29*). The position of the WD at each epoch relative to our reference frame is thus known to an accuracy that will cause only a small additional uncertainty of $\lesssim 0.5\%$ of its mass derived from the relativistic deflection of the source, even at the E6 separation of $0''.46$ (*29*). Combined with its radial velocity of $+29$ km s$^{-1}$ (*40*), the total space velocity of the Stein 2051 system is 68.8 km s$^{-1}$ with respect to the Sun.

**Relativistic deflection and mass of Stein 2051 B**

Figure 2 plots the measured source positions at the four epochs that we analyzed, showing the relativistic deflections. At each epoch they are in the direction away from the foreground WD, and by an amount inversely proportional to the angular distance between the source and the WD, as expected from Eq. 2.

We performed a model fit to the observed shifts shown in Fig. 2 using six parameters: the initial RA and Dec positions of the source, its two PM components along RA and Dec, its parallax with respect to the reference frame used, and $\theta_E$, and adopting a chi-square minimization approach (*41*). An additional constraint is that the deflection must lie along the line joining the WD and the source. Fitting this model to the 19 pairs of observed RA/Dec positions at the four useful epochs yields the predicted positions shown in Fig. 2. The observed positions are consistent with the positions expected from the model within the measurement uncertainties. The time evolution of the angular shifts is also consistent with our model (Fig. 3).

The resulting fitting parameters of our model are a source PM of $(\mu_\alpha, \mu_\delta) = (-0.4 \pm 0.05, +0.2 \pm 0.05)$ mas yr$^{-1}$, a parallax of $0.25 \pm 0.1$ mas with respect to the mean parallax of the reference stars, and an Einstein ring radius of $\theta_E = 31.53 \pm 1.20$ mas. Using Eq. 1 and the measured parallax, we find Stein 2051 B to have a mass of $0.675 \pm 0.051$ M$_\odot$.

**Astrophysics of the cool white dwarf Stein 2051 B**

Most stars end their lives as WDs—as will the Sun—and then slowly cool. Composed of degenerate matter, WDs are expected to obey a mass-radius relation (MRR) such that, as the mass of the WD increases from ~0.5 M$_\odot$ to the Chandrasekhar limit of ~1.4 M$_\odot$ (*42*), its radius decreases approximately as the inverse cube root of its mass (*43*). The MRR has a relatively large dependence on core composition, and smaller dependencies on photospheric composition, thicknesses of the H and/or He envelopes lying above the degenerate interior, and a continued small amount of shrinkage as the WD gradually cools (*44*). The vast majority of WD masses cannot be measured directly, but have to be inferred from model-dependent determinations of their surface gravities and estimates of their radii from parallax determinations and photometric or spectroscopic flux measurements (*44*), or from gravitational redshifts in cases where the true radial velocity of the WD is known from measurements of a companion star (*45*). The number of WDs whose masses and radii have been directly measured with sufficient precision to test theoretical MRRs includes just three WDs in nearby wide visual binaries [Sirius B (*46*), Procyon B (*47*) and 40 Eri B (*43*)] and about 10 WDs in short-period eclipsing binaries (*44*). However the stars in the latter group have undergone common-envelope events and have therefore not evolved in the same way as isolated single stars. The mass of a WD in a transiting binary system with an 88-day orbital period was recently measured through the photometric microlensing caused by the WD as it periodically passes in front of the G-dwarf companion (*48*).

Our direct measurement of the mass of Stein 2051 B, with an uncertainty of 0.051 M$_\odot$, provides an additional data point for comparison with theoretical MRRs and evolutionary cooling tracks. The location of Stein 2051 B in the MRR is shown in Fig. 4. We overlay a MRR for He-atmosphere, CO-core WDs (*23*) interpolated to the effective temperature of Stein 2051 B. For comparison, the MRR for zero-temperature WDs with iron cores (*49*) is also shown, but is excluded by our measurement. For a CO core, the diagram shows that our radius determination implies an expected mass of Stein 2051 B of 0.67±0.03 M$_\odot$, in agreement with our measured value of 0.675±0.051 M$_\odot$.

The position of Stein 2051 B in the theoretical Hertzsprung-Russell diagram (luminosity vs. surface effective temperature) is shown in Fig. 5 along with evolutionary cooling sequences with their cooling ages marked for CO-core WDs of masses 0.5 to 0.8 M$_\odot$ (*23*). These models have thin H ($M_H/M_{WD}$=10$^{-10}$) layers, which are appropriate for helium-atmosphere compositions. The implied WD cooling age of Stein 2051 B is 1.9±0.4 Gyr. The progenitor of the WD had a mass of 2.6±0.6 M$_\odot$, based on a recent determination of the initial-mass/final-mass relation (*50*). The corresponding theoretical pre-WD evolutionary lifetimes of these progenitors range from 0.4 to 1.3 Gyr *(51)*. Combining these with the cooling age, we find that the Stein 2051 system has a total age in the range of 1.9 to 3.6 Gyr. Unlike previous conclusions of an iron core WD, our measurement does not conflict with the age of the Universe. The derived age of the system is consistent with its moderately high space velocity, suggesting that it may be a member of the Galaxy's thick disk.

**Acknowledgments:** Based in part on observations made with the NASA/ESA Hubble Space Telescope, obtained at STScI, which is operated by the Association of Universities for Research in Astronomy, Inc., under NASA contract NAS 5-26555. HST data used in this paper are available from the Mikulski Archive for Space Telescopes at STScI (https://archive.stsci.edu/hst/search.php), under proposal ID 13457 and 14448. Support for this program was provided by NASA through a grant from STScI. We thank Jennifer Mack and Megan Sosey for help with image processing, Pier-Emmanuel Tremblay for useful discussions, Armin Rest for assistance with the Pan-STARRS data, Denise Taylor for support with the observations, and Greg Bacon for help with the preparation of an animated movie. The Pan-STARRS1 Surveys (PS1) have been made possible through contributions of the Institute for Astronomy, the University of Hawaii, the Pan-STARRS Project Office, the Max-Planck Society and its participating institutes, The Johns Hopkins University, Durham University, the University of Edinburgh, Queen's University Belfast, the Harvard-Smithsonian Center for Astrophysics, the Las Cumbres Observatory Global Telescope Network Incorporated, the National Central University of Taiwan, the Space Telescope Science Institute, NASA Administration under Grant No. NNX08AR22G, the National Science Foundation under Grant No. AST-1238877, the University of Maryland, and Eotvos Lorand University. MD thanks Qatar National Research Fund (QNRF) for support by grant NPRP 09-476-1-078.

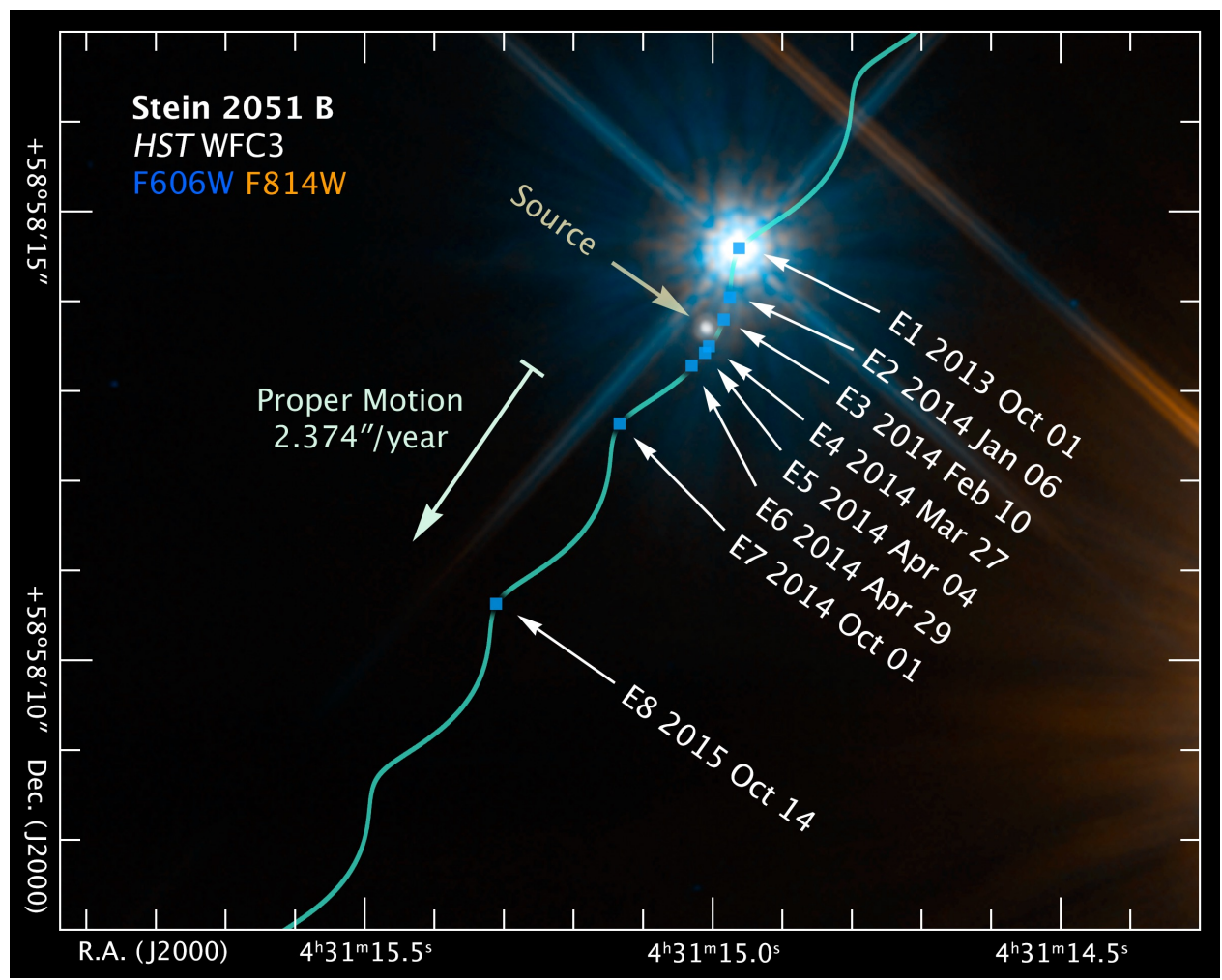

**Fig. 1.** ***Hubble Space Telescope*** **image showing the close passage of the nearby white dwarf Stein 2051 B in front of a distant source star.** This color image was made by combining the F814W (orange) and F606W (blue) frames, obtained at epoch E1. The path of Stein 2051 B across the field due to its proper motion towards south-east combined with its parallax due to the motion the Earth around the Sun, is shown by the wavy cyan line. The small blue squares mark the position of Stein 2051 B at each of our eight observing epochs, E1 through E8. Its proper motion in one year is shown by an arrow. Labels give the observation date at each epoch. The source is also labeled; the motion of the source is too small to be visible on this scale. Linear features are diffraction spikes from Stein 2051 B and the red dwarf star Stein 2051 A, which falls outside the lower right of the image. Stein 2051 B passed 0.103 arcsec from the source star on 5 March 2014. Individual images taken at all the 8 epochs, and an animated video showing the images at all epochs are shown in Fig S1 and Movie M1 (*29*).

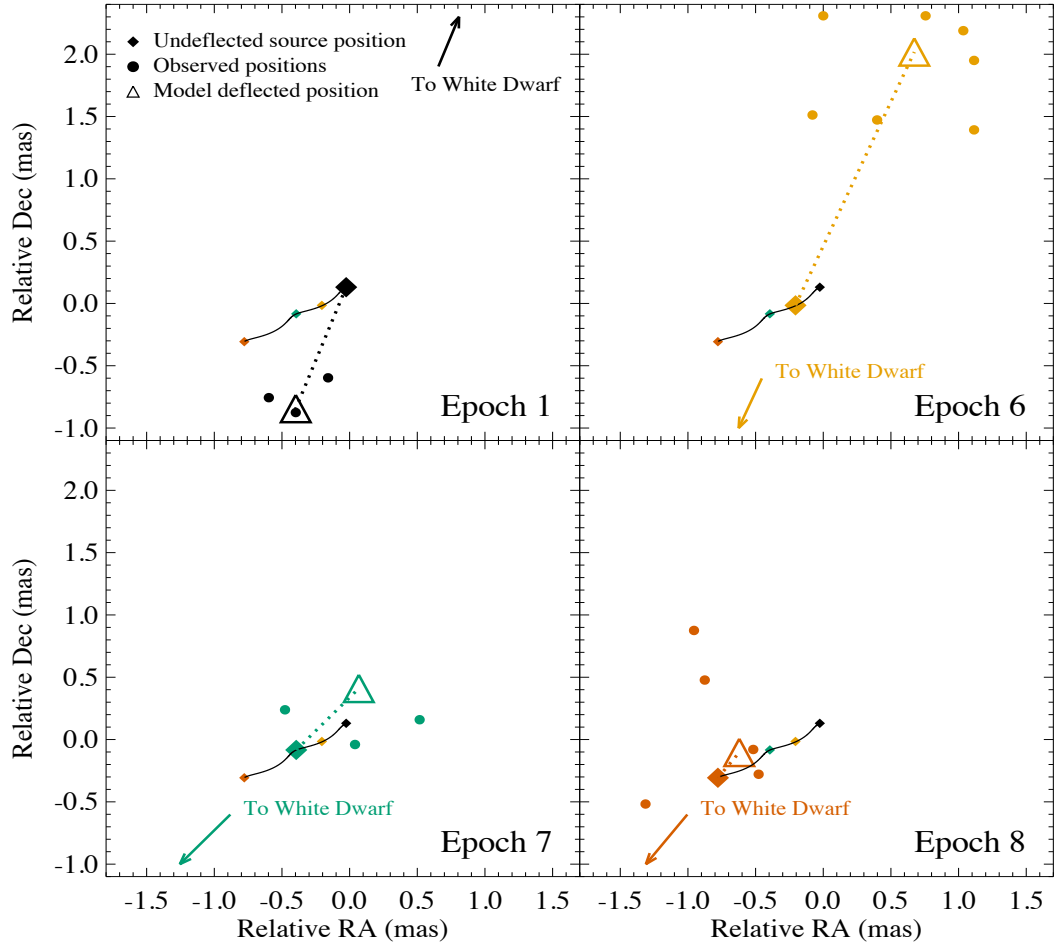

**Fig. 2.** *Hubble Space Telescope* **measurements of the background star's positions at epochs 1, 6, 7 and 8**. The solid dots are the observed positions of the source for each exposure, color-coded for each epoch. The origin corresponds to the undeflected source position at E1, and the Relative RA in the x-axis corresponds to -ΔRA×cos (Dec). The undeflected positions of the source are plotted as solid diamonds, connected with a solid line showing its small parallax and slow proper motion to the southeast. Solid arrows indicate the direction toward the white dwarf at each epoch. At each epoch, the source position is seen to be deflected from its undeflected location, along the direction away from the white dwarf. We modeled the measurements with an Einstein ring of radius 31.53 mas. The model predicted deflected positions are shown as open triangles, which are joined to the undeflected source positions by dotted lines.

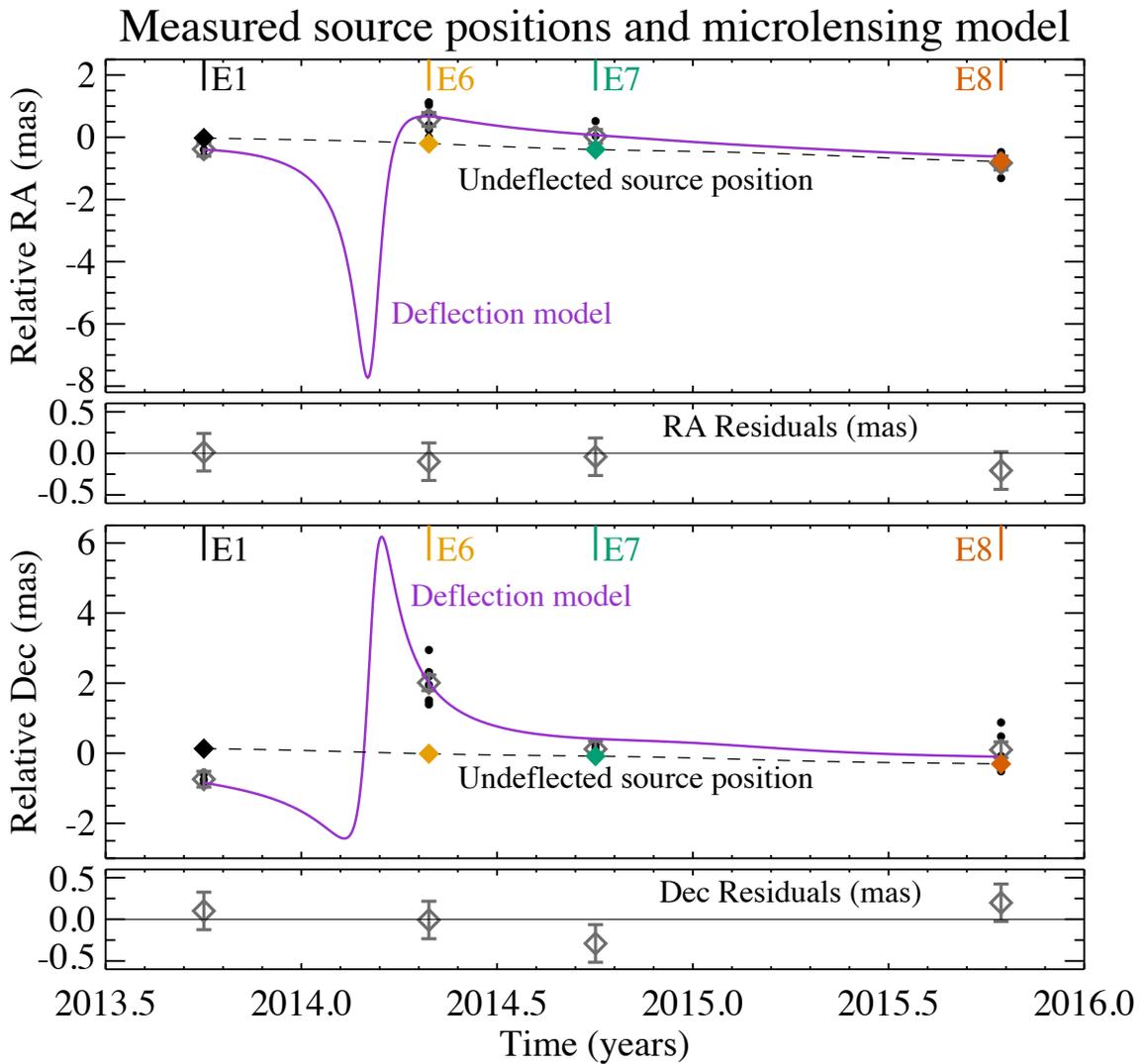

**Fig. 3. Measured and model undeflected RA and Dec positions of the background source as a function of time.** The positions are relative to the undeflected source position at E1. Solid diamonds show the undeflected positions, color-coded with the same colors as in Fig.2, and connected by dashed lines showing the small parallax and proper motion. The measured deflected positions are plotted as filled black circles, their mean at each epoch is shown as a diamond along with the standard deviations of the mean. Our model fit, with an Einstein ring radius of 31.53 mas, is shown as a solid purple curve. The RA and Dec residuals after subtracting the model from the mean observed positions are shown in separate panels.

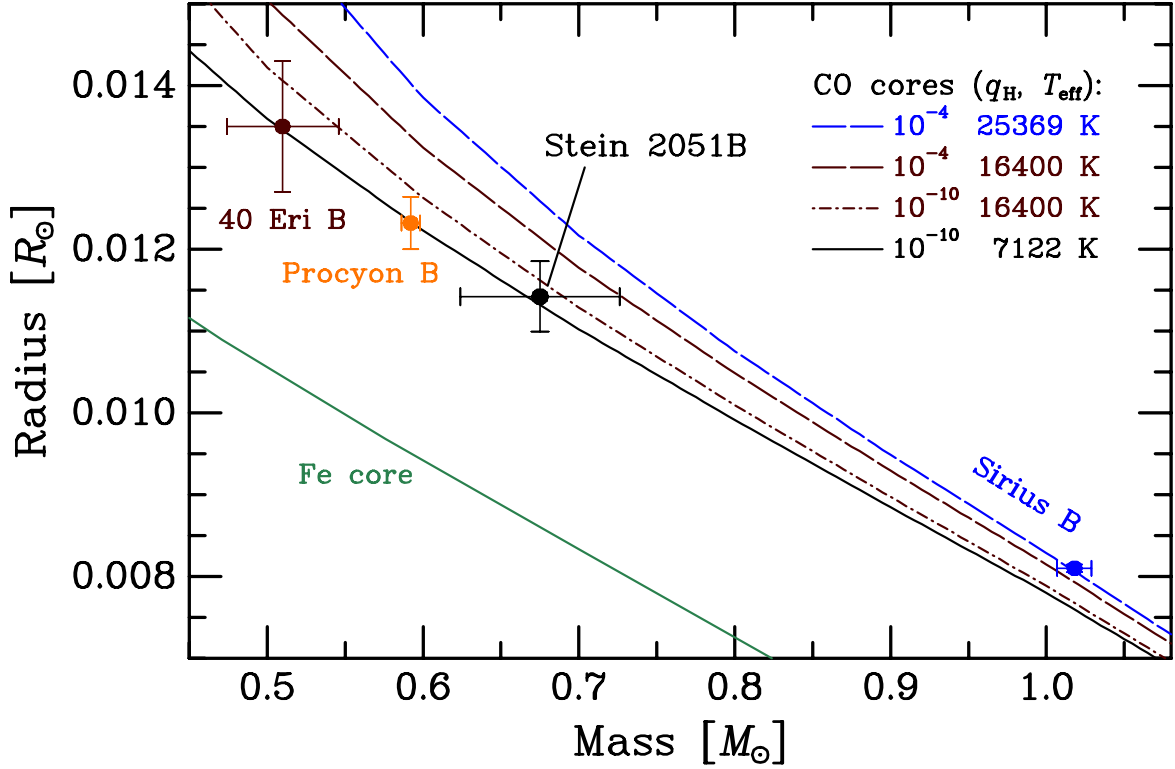

**Fig. 4. Mass-radius diagram for Stein 2051 B and three nearby white dwarfs in visual binaries.** Data points with error bars show the masses and radii for Stein 2051 B (this paper; black), 40 Eridani B (*43*; dark brown), Sirius B (*46*; blue), and Procyon B (*47*; orange). The black curve plots a theoretical mass-radius relation (*23*) for carbon-oxygen core white dwarfs with the parameters of Stein 2051 B (thin hydrogen layer, $q_H = M_H/M_{WD} = 10^{-10}$; effective temperature 7122 K). This curve is also appropriate for the similar white dwarf Procyon B. The blue curve shows the relation (*23*) for thick hydrogen-layer CO white dwarfs with the effective temperature of Sirius B, and the dark brown curves plot the relations for thick and thin hydrogen-layer CO white dwarfs with the temperature of 40 Eri B. The green curve shows the theoretical relation for zero-temperature white dwarfs with iron cores (*48*). The mass of Stein 2051 B inferred from the astrometric microlensing, $0.675 \pm 0.051$ M$_\odot$, is consistent with the CO core expected from normal stellar evolution.

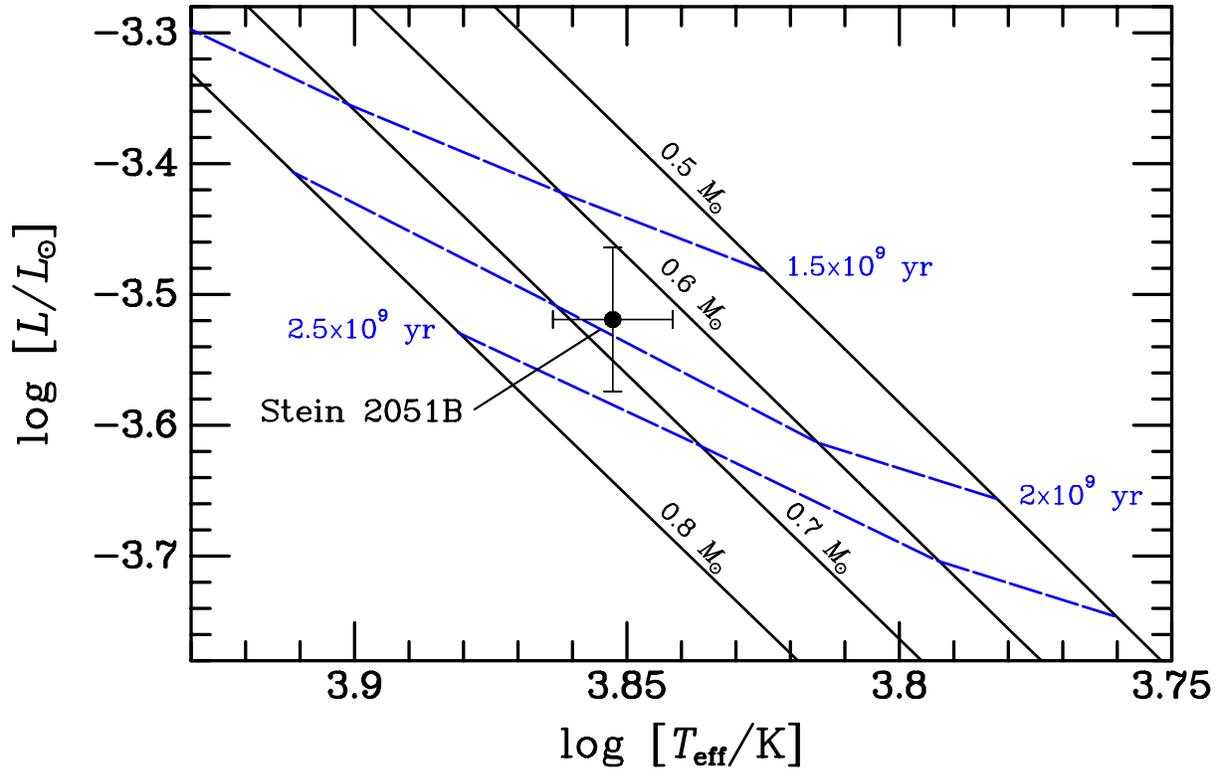

**Fig. 5.**. **Theoretical WD cooling tracks (*23*).** Cooling tracks are shown for four masses (solid lines), along with isochrones showing the WD cooling ages (dashed lines). The position of Stein 2051 B agrees within the uncertainties with that expected for our measured mass. The implied cooling age of Stein 2051 B is 1.9±0.4 Gyr.

**Table 1: Details of the HST Observations.** The numbers in the "No. of exp" column correspond to the number of exposures taken with the corresponding "Exp time" specified in the previous column. The last column gives the projected separation between the lens and the undeflected position of the source.

| Epoch | Obs. date | Filter | Subarray size (pixels) | Exp. time (s) | No. of exp. | Lens-source separation ($''$) |
|---|---|---|---|---|---|---|
| 1 | 1 Oct 2013 | F814W | 2048×2048 | 0.5/250 | 2/3 | 0.943 |
|   |   | F606W | 2048×2048 | 0.5/250 | 2/2 |   |
| 2 | 6 Jan 2014 | F814W | 2048×2048 | 1/270 | 2/3 | 0.434 |
|   |   | F606W | 2048×2048 | 0.5/270 | 2/2 |   |
| 3 | 10 Feb 2014 | F814W | 1024×1024 | 1/50 | 3/3 | 0.203 |
|   |   | ″ | 2048×2048 | 1/50 | 1/1 |   |
|   |   | F763M | 512×512 | 2/90 | 11/11 |   |
| 4 | 27 Mar 2014 | F814W | 1024×1024 | 1/50 | 3/3 | 0.205 |
|   |   | ″ | 2048×2048 | 1/50 | 1/1 |   |
|   |   | F763M | 512×512 | 0.5/8 | 11/19 |   |
| 5 | 4 Apr 2014 | F814W | 1024×1024 | 1/50 | 3/3 | 0.262 |
|   |   | ″ | 2048×2048 | 1/50 | 1/1 |   |
|   |   | F763M | 512×512 | 0.5/8 | 11/19 |   |
| 6 | 29 Apr 2014 | F814W | 1024×1024 | 2/3/75 | 8/8/8 | 0.462 |
| 7 | 1 Oct 2014 | F814W | 2048×2048 | 1/240 | 3/3 | 1.478 |
|   |   | F606W | 2048×2048 | 0.5/240 | 2/2 |   |
| 8 | 14 Oct 2015 | F814W | 2048×2048 | 1/240 | 3/5 | 3.897 |
|   |   | F763M | 2048×2048 | 0.5 | 2 |   |

# Supplementary Materials for

Relativistic deflection of background starlight measures mass of a nearby white dwarf star

Kailash C. Sahu, Jay Anderson, Stefano Casertano, Howard E. Bond, Pierre Bergeron, Edmund P. Nelan, Laurent Pueyo, Thomas M. Brown, Andrea Bellini, Zolt Levay, Joshua Sokol, Martin Dominik, Annalisa Calamida, Noé Kains, Mario Livio

Correspondence to: ksahu@stsci.edu

**This PDF file includes:**

Materials and Methods
Supplementary Text
Table S1
Figs. S1 – S10
References (52 – 57)

**Other Supplementary Materials for this manuscript include** Supporting Movies Available from STScI website: http://hubblesite.org/news_release/news/2017-25

**Materials and Methods**

<u>Details of the *Hubble Space Telescope* Observations</u>
The observations were summarized in Table 1. We used the Wide Field Camera 3 (WFC3), which provides a plate scale of 39.6 mas pixel$^{-1}$. At each epoch, we employed the broad-band F814W (roughly equivalent to *I*-band) filter. This was supplemented with broad-band F606W (wider version of *V*) at several epochs, and with the medium-band F763M filter at the epochs of closest approach to the background star. Numbers in the filter names indicate their central wavelength in nanometers. Exposures were taken at several pointings dithered by ~2–5″. Fig. S1 shows the central ~15×12″ region of the field as observed at different epochs in F814W filter. The images taken at our 8 observed epochs clearly show the motion of Stein 2051 B.

Our observing strategy was adjusted at each epoch, taking the separation between the WD and the background star into account. At large separations, we used a 2048×2048 pixel subarray, which included Stein 2051 A and B, a grid of 26 fainter background reference stars (visual magnitudes ~17–21), and the source star. At these epochs, we obtained both long exposures in which the reference stars and the source were well-exposed but A and B were saturated, and very short ones at the same telescope pointing in which B was not saturated and its location could be measured precisely. For observing efficiency with the more numerous shorter exposure times, we used 512×512 and 1024×1024 pixel subarrays in order to avoid interrupting the observations for data-buffer dumps.

<u>Data analysis procedure</u>
The raw WFC3 images were corrected for bias and flat field via the standard Space Telescope Science Institute (STScI) pipeline reductions (*30*). Correction for charge-transfer inefficiency was done using publicly available software from STScI (*52*). The resulting images were then used for photometric and astrometric analysis.

We used empirical effective point-spread functions (PSFs) available from STScI (*53*) and a centroiding algorithm (*31*) to measure the positions of the stars at all epochs. All the stars in the long exposure ($T_{exp} > 75$ sec) images are well measured with >10,000 electrons, for which the expected centroiding accuracy is typically 0.01 to 0.02 pixels (*54*). (The position measurements of the WD and the source needed special treatment, which is described in the next 2 sections). We then used the distortion corrections (*32*) to convert the measured positions to an undistorted frame. The positions of 26 stars closest to Stein 2051 B at epoch E1 (shown in Fig. S2) were used as our reference frame. The positions of all the stars measured at other epochs were transformed (offset, rotation, skew, and scale) to this reference frame using the procedure described below.

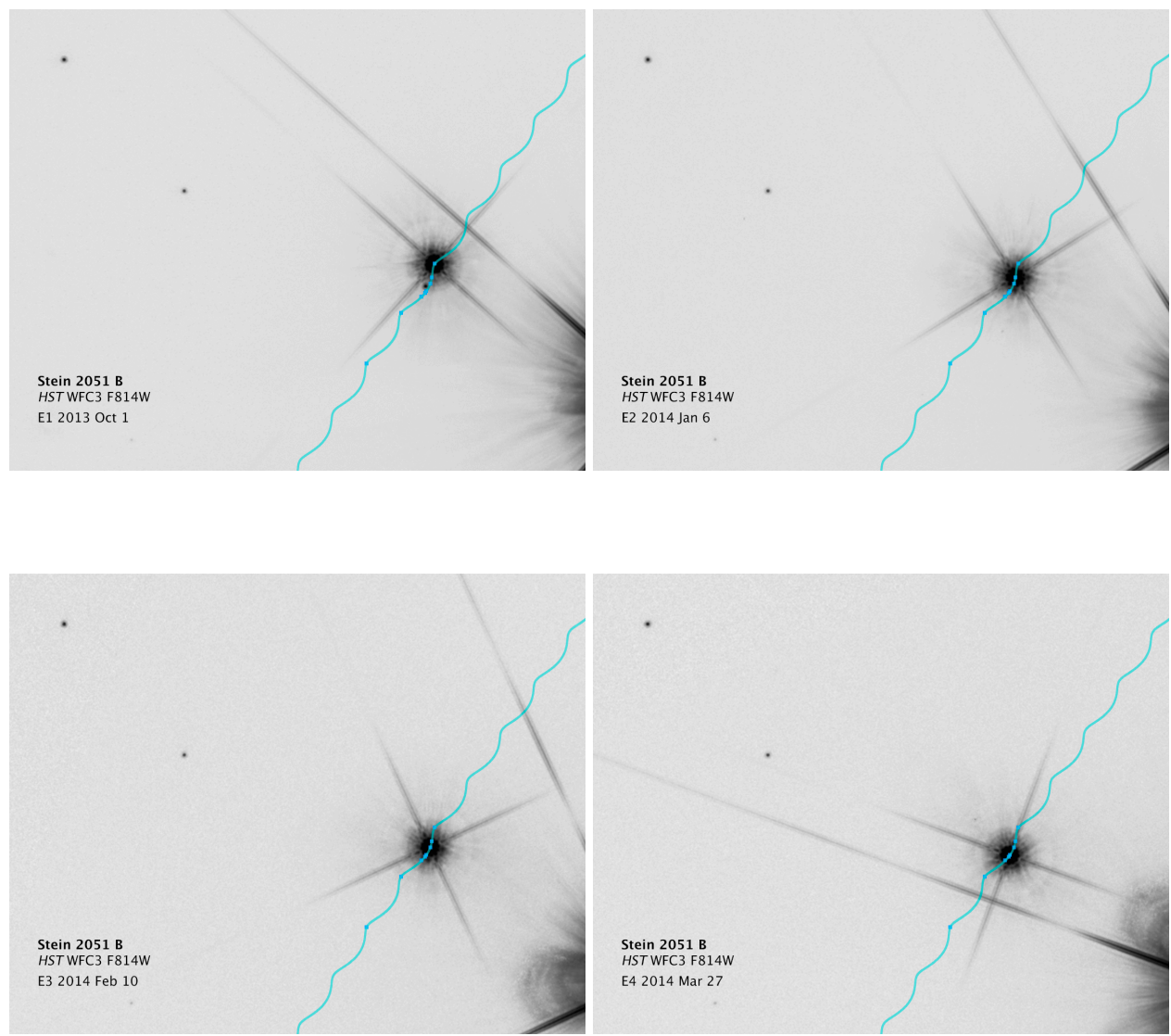

**Fig. S1. HST/WFC3 images of the region around Stein 2051 B at different epochs.** North is up, and east is to the left. The size of the image is ~15×12 arcsec. The images include two background stars for reference. Stein 2051 A is located just outside the frames, at the bottom right, and its diffraction spikes can be seen crossing the images. Stein 2051B moves toward the southeast along the wavy blue line, due to proper motion and parallax. The expected positions of Stein 2051 B at all the observation epochs are shown as small diamonds. The observation epochs and the corresponding dates are specified at the bottom left of each panel. An animated video (Movie M1) shows the images at all the observed epochs.

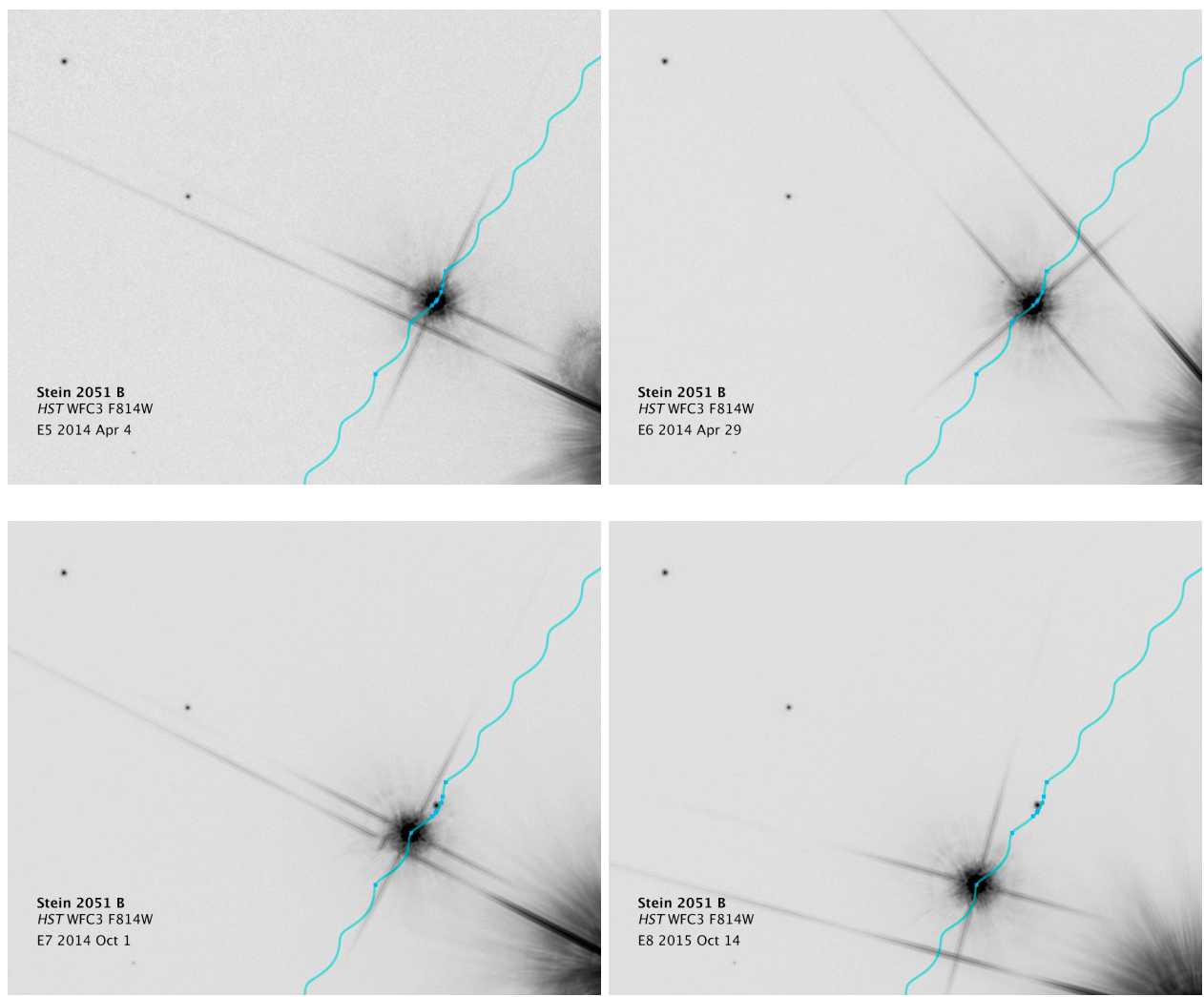

**Fig. S1. continued.**

Since the positions of the individual stars are expected to change due to their PMs, we first needed to determine their PMs, for which we employed an iterative procedure. We determined the positions of all the reference stars at all epochs in the reference frame described above. We carried out a linear fit to these positions to derive a PM for each star. The derived PMs were then used to determine the revised positions of all stars at all epochs. We iterated this procedure of using the revised positions and PMs of the stars to further revise their positions and PMs. We rejected the highest sigma point after each complete iteration. We repeated this procedure until the highest sigma point was no more than a preset value ($6\sigma$ in this case). Only 6 points were rejected by this procedure, all of which were affected by cosmic ray hits on the detector.

Figure S3 shows the derived proper motions of the 26 reference stars. Fig. S4 shows the PM errors of the 26 reference stars, the median of which is 0.26 mas/yr. We used the derived PMs to calculate the model positions of the reference stars at all epochs. The model positions of the reference stars at different epochs then form the reference frames for the respective epochs. Fig. S5 shows a histogram of all residual PMs for all the stars. We note that, some of the reference stars are brighter than the source, and some are fainter. Since our exposures were optimized for the source star, the fainter stars have lower positional accuracy. Our algorithm gives appropriately lower weights to the fainter stars in deriving the transformations for the measured positions. Fig. S6 shows the number of observed points as a function of normalized residuals, and the expected Gaussian distribution (the red curve). The number distribution of the observed points is consistent with the expected Gaussian distribution. As shown by previous similar studies, the final reference-frame positions are expected to be internally accurate to better than 0.01 pixel *(55, 56)*.

We derived the transformation parameters needed to convert the measured positions to the model positions using our software developed for this purpose, giving appropriate weights to individual measurements *(55)*. We then used these transformation parameters to determine the positions of the WD and the source in this reference frame.

It is clear from the above procedure that all our measurements are with respect to the average motion of the 26 reference stars. In the procedure described above, we specifically solved for the PMs of the reference stars, but we ignored their parallaxes. The reason for adopting this approach is described below.

We used the photometric magnitudes and colors determined from our *HST* data and from the Pan-STARRS *(33)* and 2MASS databases *(34)* to estimate the distances from Earth to the reference stars and the source. We assumed that they are dwarfs (giants would generally be at implausibly large distances of >50 kpc), and used theoretical isochrones *(35)* to calculate their distances. The reference stars are found to lie at distances of ~0.8–2.1 kpc, corresponding to parallaxes of 0.5 to 1.25 mas. The mean of the estimated parallaxes of the reference stars is $0.8 \pm 0.2$ mas. The source itself is estimated to be a K dwarf at 2.0 kpc. Thus the parallactic motion of the reference stars and the source with respect to the mean parallax is less than or comparable to our individual measurement uncertainty of 0.4 mas. So, instead of fitting for the small parallaxes of every reference star, we ignored the parallax of the reference stars in the initial transformation process, but included the parallax of the source as a model parameter in fitting positions at a later stage.

It is worth pointing out here that the small uncertainty in the parallax of the source has little effect on our final mass measurement. First, as seen from Eq. 1, the size of $\theta_E$ depends on the reduced parallax, which can be expressed as the parallax of the lens minus the parallax of the source. Thus our reference frame can have any parallax as long as we measure the lens and source in the same reference frame. In addition, the distance to the lens (5.52 pc, corresponding to a parallax of ~181 mas) is much smaller compared to the distance to the source (2 kpc, corresponding to a parallax of 0.5 mas). We can thus assume $1/D_r = (1/D_l - 1/D_s) \sim 1/D_l$. In our specific case, the error in the mass measurement would be less than ~1% even if we ignore the source parallax by assuming it to be zero (but we do fit for the parallax of the source in our analysis).

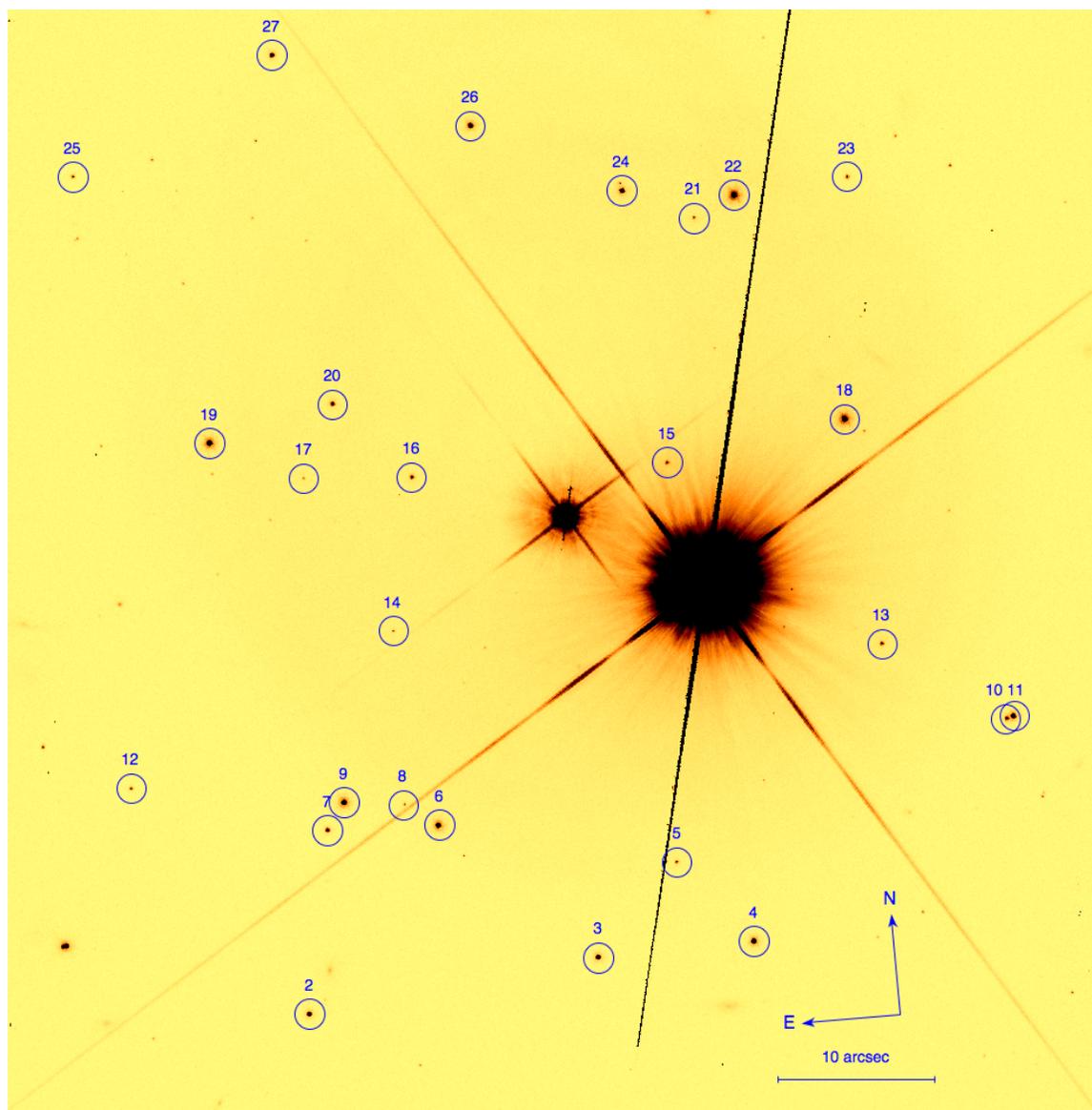

**Fig. S2. WFC3/HST image of the field in F814W filter, showing the reference stars.** Stein 2051 A is the brightest star in the field, and Stein 2051 B is the second brightest.

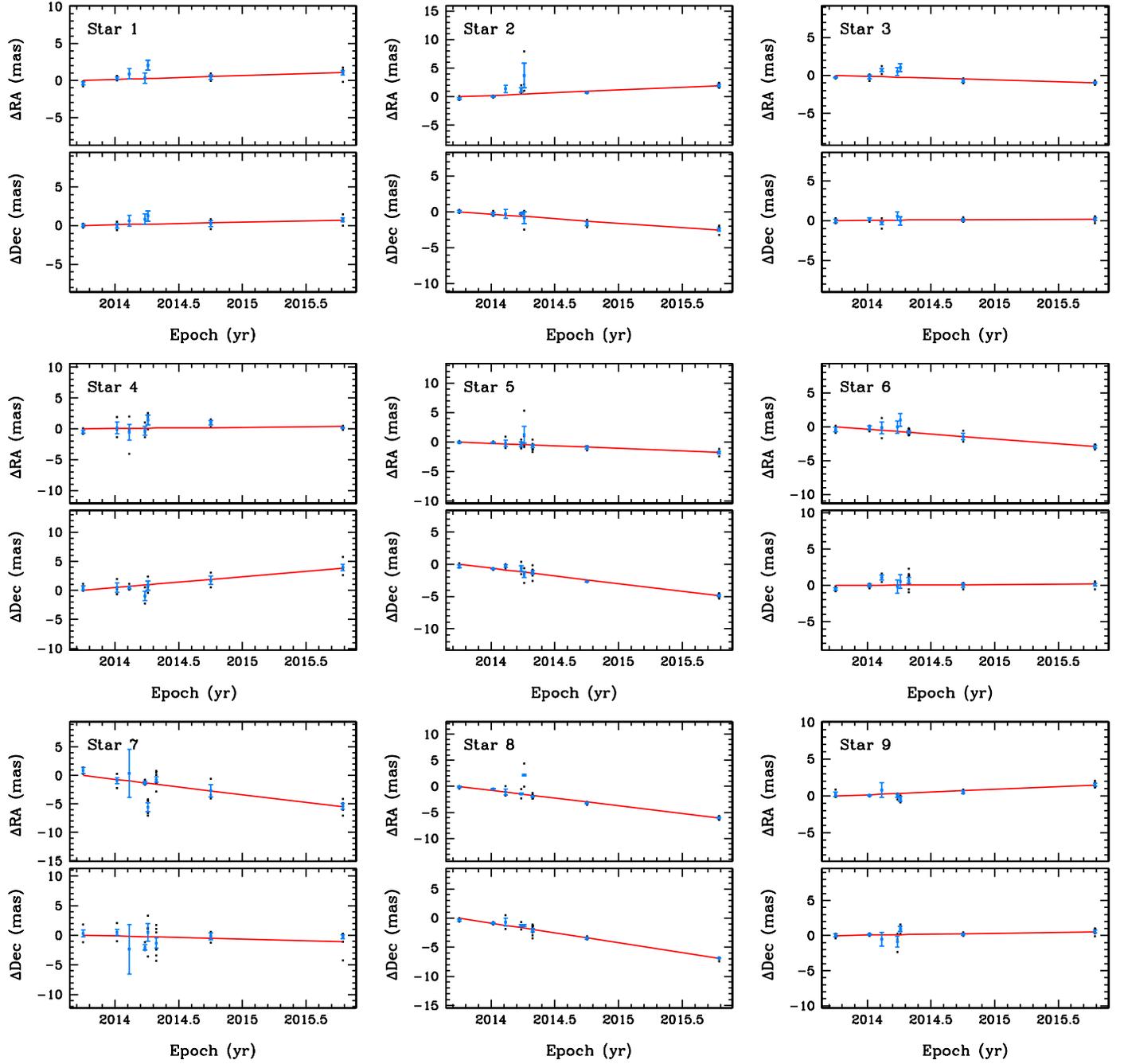

**Fig. S3. Proper motions of the reference stars along X (RA) and Y (Dec) axes.** The proper motions shown in the RA axis correspond to -ΔRA×cos (Dec).

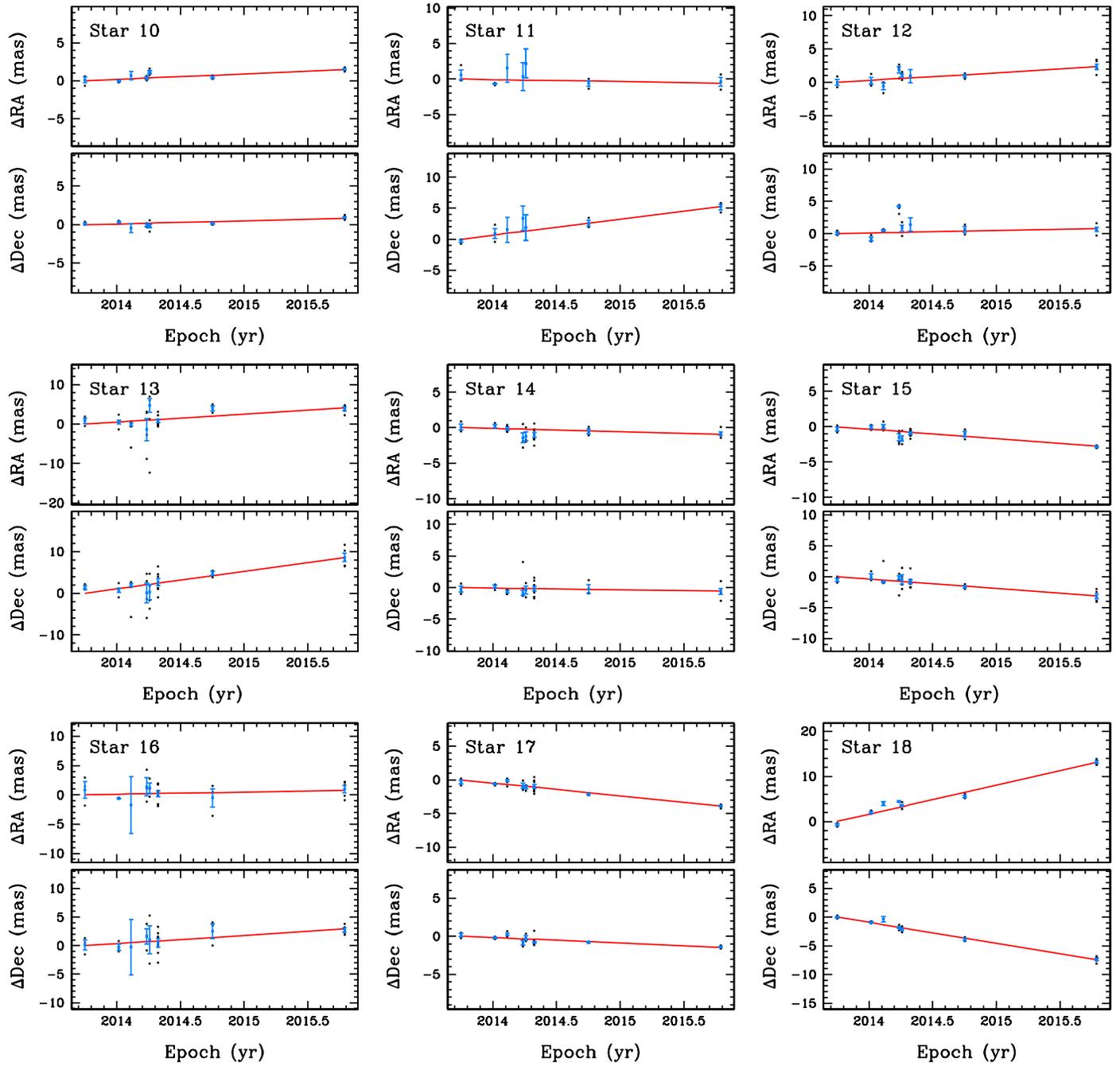

**Fig. S3. continued.**

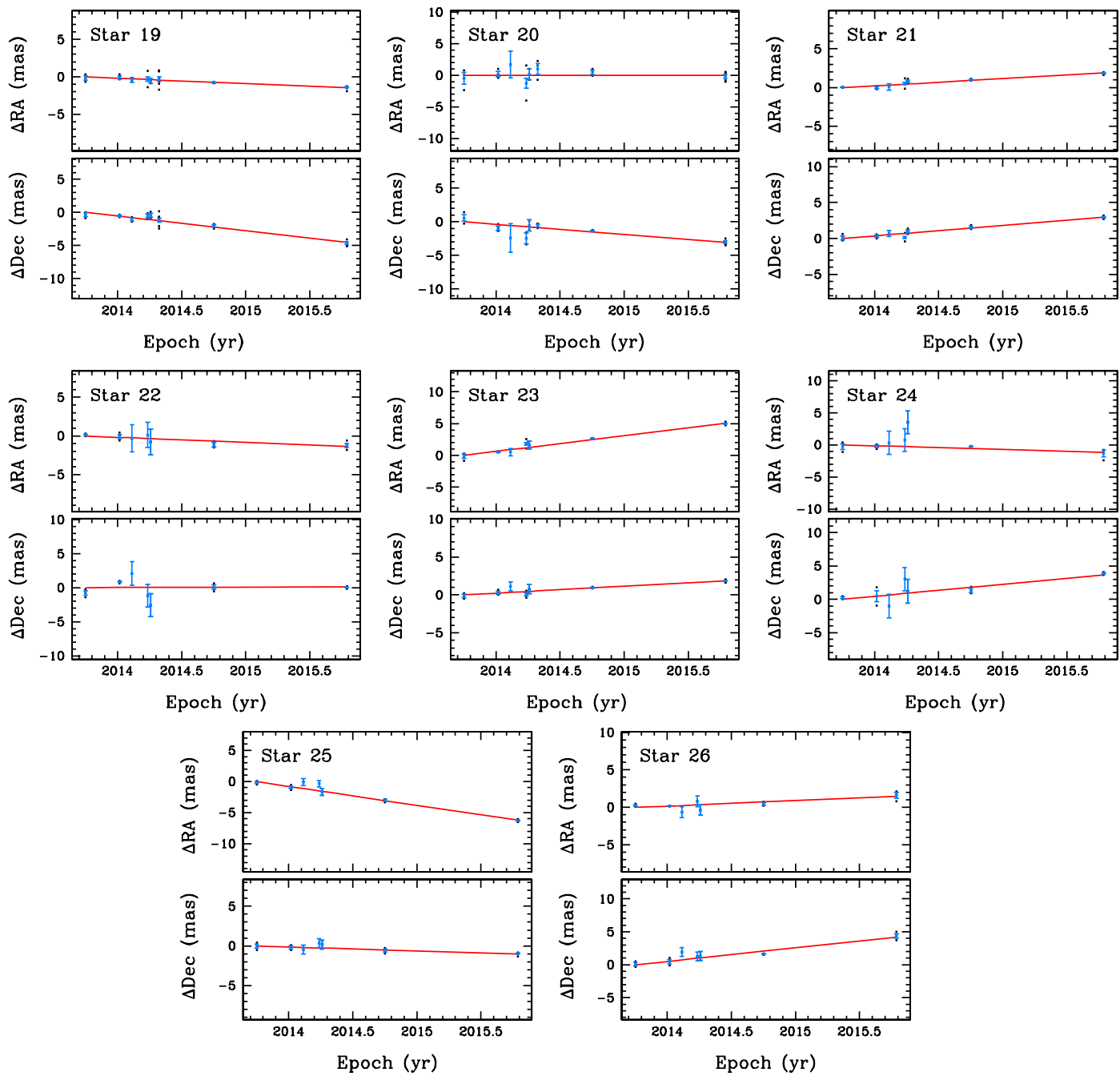

**Fig. S3.** continued.

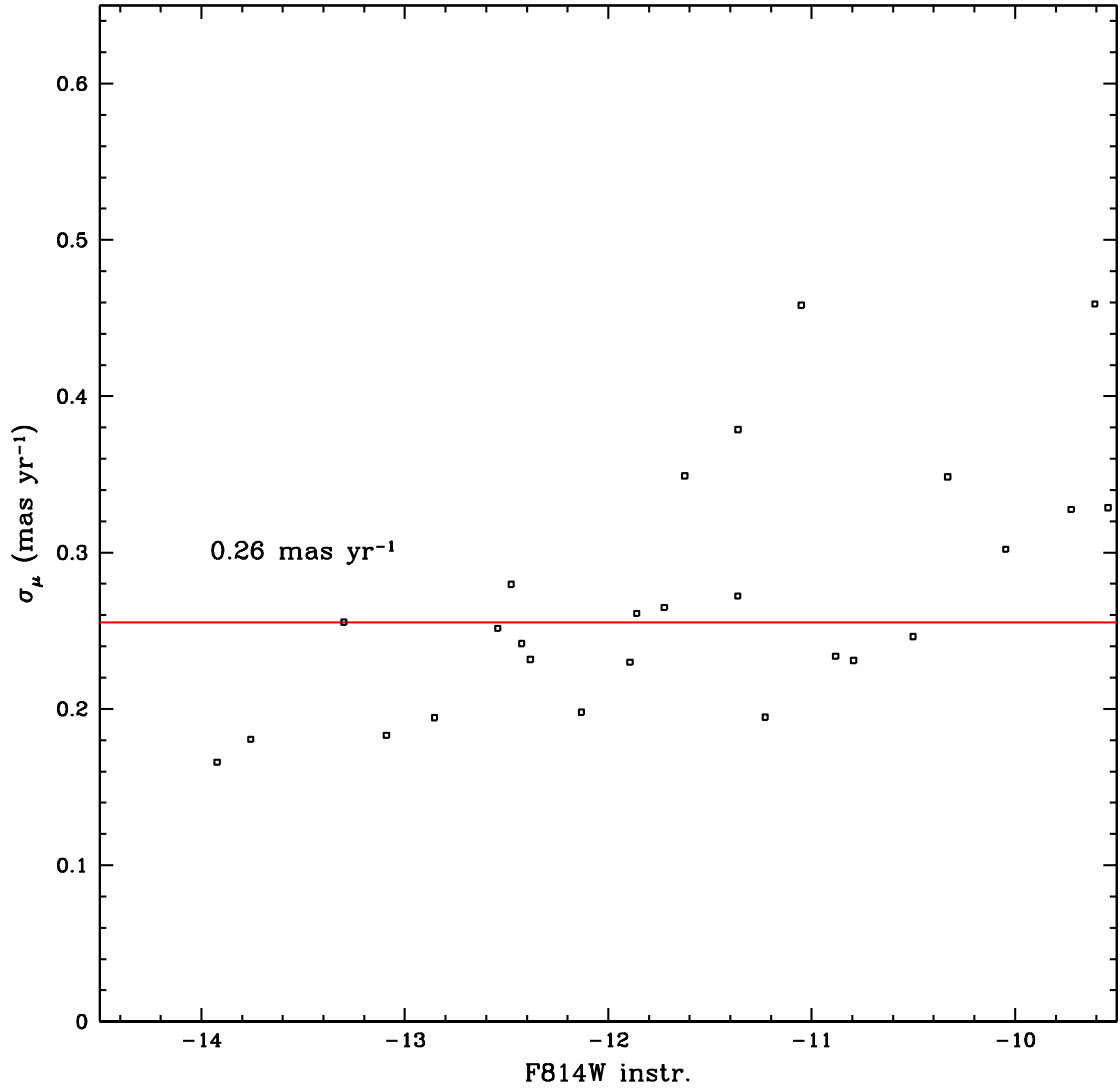

**Fig. S4. Proper motion errors of the reference stars.** The X-axis shows the instrumental magnitude of the star, defined as -2.5*log (flux in electrons). The Y-axis shows the PM error defined as $\sigma_\mu = \left[ (\sigma_{\mu x}^2 + \sigma_{\mu y}^2)/2 \right]^{0.5}$ where $\sigma_{\mu x}$ and $\sigma_{\mu y}$ are the PM errors along x and y axes.

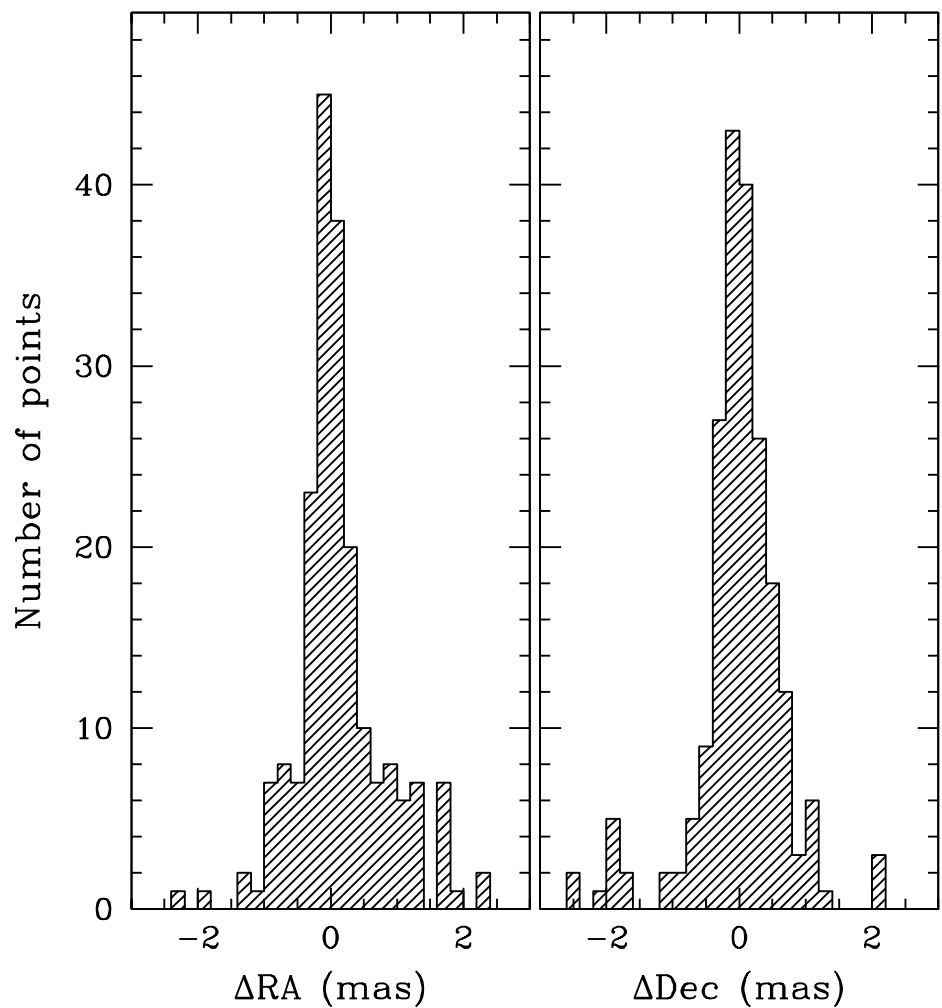

**Fig. S5. Histogram of the PM residuals for all the reference stars in RA and Dec.** Some reference stars are brighter than the source, and some are fainter. The residuals are higher for fainter stars as expected; our analysis procedure gives appropriately lower weight to the fainter stars.

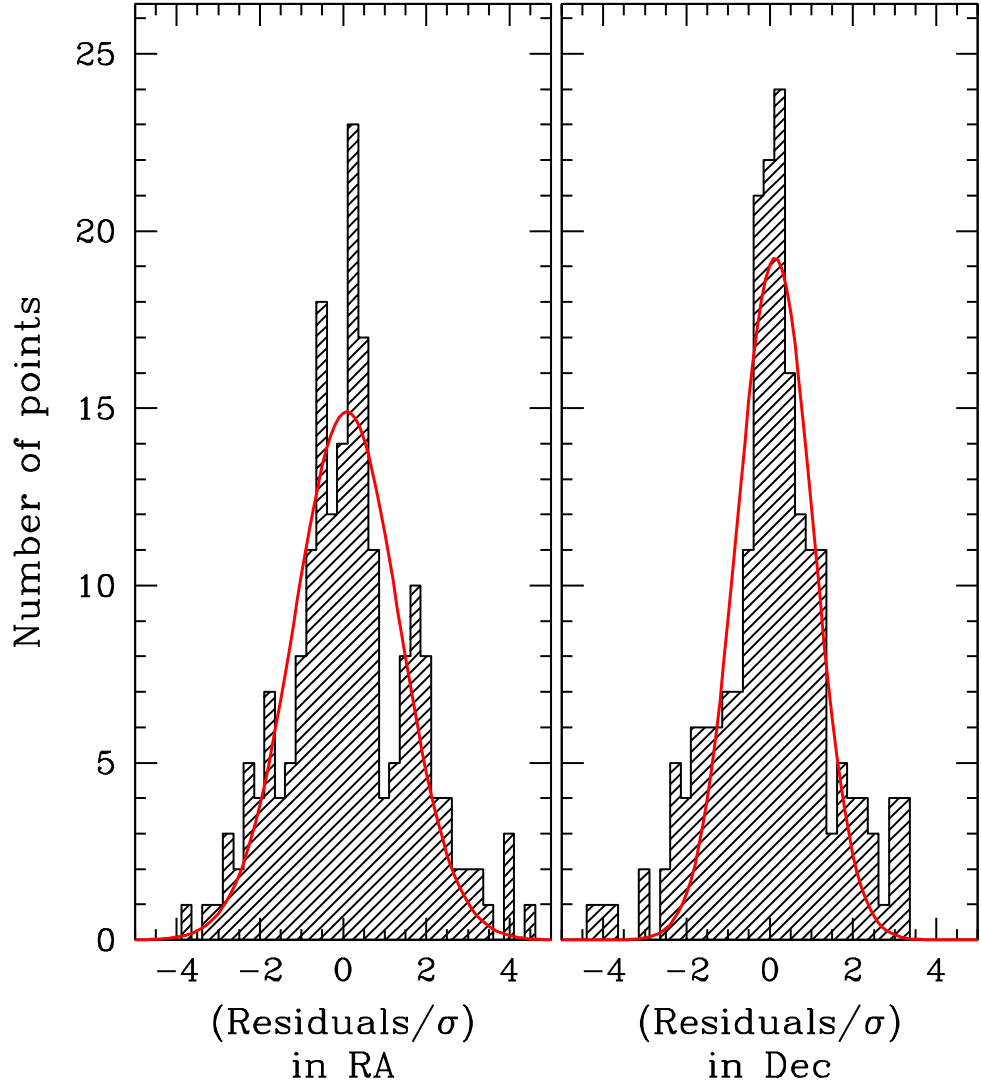

**Fig. S6. Histogram of the normalized residuals in RA and Dec.** The number of observed points is shown as a function of the PM residuals divided by the σ of the residuals. The distribution of the observed points is consistent with the expected Gaussian distribution shown by the red curve.

Measurements of the WD positions

The position of Stein 2051 B could only be determined in the short-exposure frames, as it was saturated in the long exposures. To correct for the (small) telescope pointing drift between the short and long exposures, we used a small subset of bright reference stars detected in both exposures to align the short exposures with the long ones, thus deriving the position of the WD in the long-exposure frames. This generates an extra uncertainty in the position of the WD, with a resulting uncertainty of ~2 mas. However, it is easy to show from Eq. 1 and 2 that for u >> *1* (as it is in our case), the deflection δθ can be expressed as (15)

$$\delta\theta \sim \theta_E{}^2/\Delta\theta.$$

We also note from Eq. 1 that $\theta_E{}^2$ is directly proportional to M, the mass of the lens. The positional uncertainty of 2 mas of the WD corresponds to ~0.5% of even the closest separation between the WD and the source ($\Delta\theta$), which is 0″.46 at E6. The observed uncertainty in the position of the WD thus corresponds to an uncertainty of no more than 0.5% in the mass of the lens, and 0.25% in $\theta_E$.

Measurement of source positions

The radial profile of the WFC3 PSF in F814W is shown in Fig. S7, which will help appreciate the issues involved in measuring the position of the faint source close to the WD which is ~400 times brighter. As we see from Table 1, the distance of the target from the WD at these 4 epochs are:
E1: 23.8 pixels
E6: 11.7 pixels
E7: 37.3 pixels
E8: 98.4 pixels

The radial profile of the PSF shows that the contribution of the WD beyond 12 pixels is negligible. This suggests that the contribution from the WD at position of the target is negligible at epochs E1, E7 and E8. Indeed, subtraction of the PSF at these epochs has no effect on the position measurements of the source. So our standard method of using a library PSF for measuring the positions worked well for epochs E1, E7 and E8.

At epochs E3, E4, and E5, the WD was very close to the source ($\Delta\theta \simeq$ 0″.20–0″.26), and the measurements of the source had large uncertainties (~0.1 pixels) even after PSF subtraction. In epoch E2 frames, the source lay on a diffraction spike of the WD, and consequently the measurements had large uncertainties of ~0.1 pixels. Thus the observations at these four epochs were useful in refining the parallax and PM of the WD, but were not useful for the deflection measurements of the source.

Analysis of E6 observations

Using the radial profile of the PSF, we estimate that the contribution of the WD in E6 would be ~100 electrons at the position of the target, while the target itself is expected to have a peak of about 2000 electrons. Since we are aiming for high astrometric accuracy, it is important to carry out a proper PSF subtraction of the WD before measuring the position of the source.

We first attempted to use the M dwarf Stein 2051 A as a reference PSF. Our deflection measurements of the source were carried out in the I-band, since that would maximize the source-to-WD flux ratio, and maximize the signal we are looking for relative to the noise (after the PSF subtraction).

However, the task of using Stein 2051 A as the reference PSF proved to be challenging since Stein 2051 A is red and Stein 2051 B is blue. When we subtracted the WD using the PSF derived from Stein 2051 A, we found this PSF to be too broad to properly fit the WD. Since the size of the PSF is expected to scale with the effective wavelength of observation, we then scaled this PSF by 0.97 (which is the ratio of the effective wavelengths of B to A) and subtracted from the WD. The residuals were much less prominent after this subtraction. However, this procedure of subtracting the PSF only works on a stack, which would provide only a single measurement, rather than multiple independent observations that could be used to give us a sense of the uncertainties in our measurements. Since E6 measurements are the most important ones in our analysis, we decided to seek a more robust measurement approach.

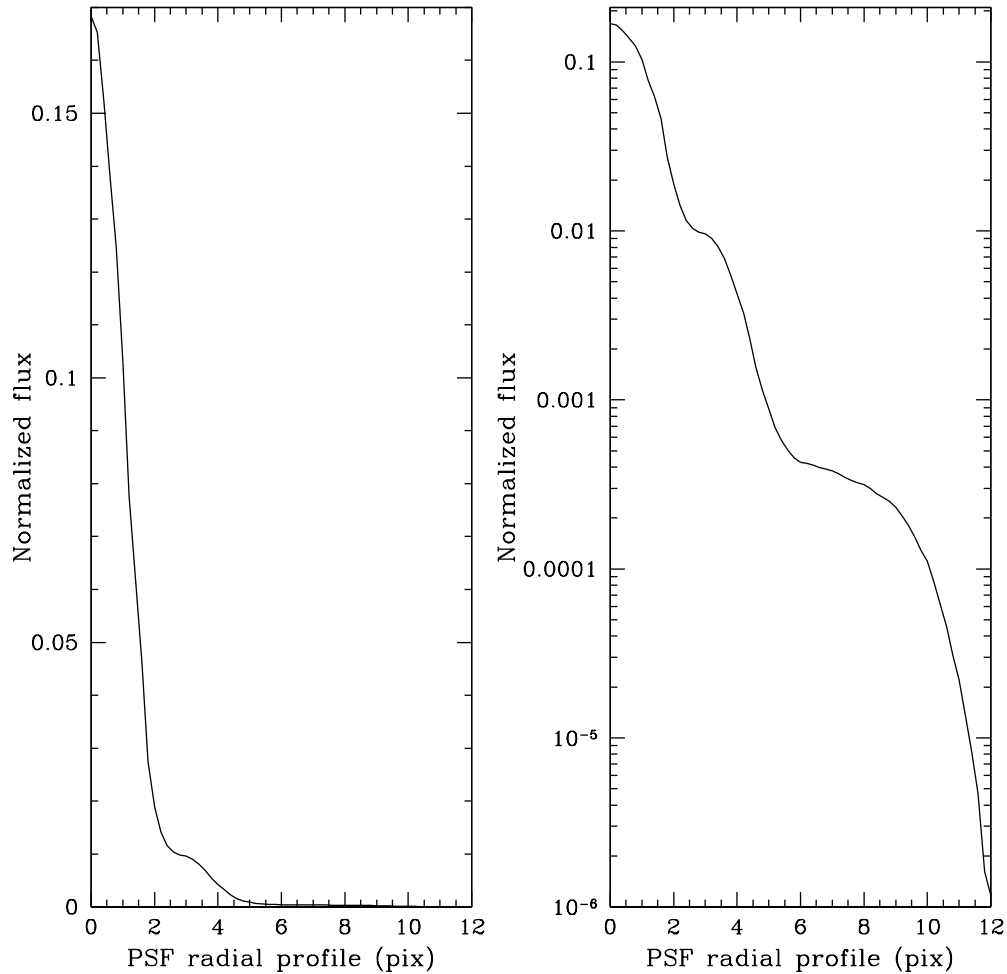

**Fig. S7: Radial profile of the WFC3 PSF in F814W filter.** The profile was made from a large number of observations, with the stars located at different parts of the detector.

In order to both get a better measurement *and* a confidence range (which comes from multiple independent observations), we set out to subtract the WD differently — in each independent exposure. We gathered all the observations of the WD during all of the epochs we had available. There were 29 observations in total, 8 of which were taken in E6.

These observations were stored in a data cube of 29 raster images. Each raster image has 101x101 pixels, with the WD centered at (51,51), the center of the raster. Each of these rasters contains the image of the WD in F814W, along with the source at some location relative to the WD. HST is known to have a breathing pattern which causes a slight variation of the PSF even during the same orbit, so the PSFs are expected to be slightly different for each exposure, although the effect at the outer parts (> 10 pixels) is expected to be small. Of course, every observation will be at a slightly different x/y pixel phase compared to that of the exposure where we are hoping to subtract the star. But the sample of 29 images is large enough that there should be at least a few observations where the WD is centered such that we can get an accurate subtraction.

We took every non-E6 WD observation, scaled it, and subtracted it from every E6 WD observation, resulting in a series of images. The result is shown in the upper panel of Fig. S8. Along the top row, we show the WD+target (with the target at a different location with respect to the WD) for the 21 F814W exposures in E1 through E7, excluding E6. In column 1, each E6 observation of the WD is in a different row, and in the columns down the row, we show the subtraction by the different comparison images, taken from Row 1. The 5×5 pixels in the vicinity of the target were not used in the subtraction scaling.

The lower panel of Figure S8 shows a close-up of the top two E6 observations and the first six subtractions. The source is at about 10 o'clock position relative to the WD. Some comparison images clearly provide cleaner subtractions than others.

Thus, for each of the eight E6 observations we have 21 WD-subtracted rasters. In each one, we measure the position and flux of the target star using our standard WFC3 F814W PSF. We use a 3×3-pixel box (centered on the target's brightest pixel) and measure a position, flux and sky. We also measure the linear correlation coefficient ($r$) to see which observations are most consistent with a PSF shape, and thus presumably least contaminated by poor WD subtraction.

This process gives us a list of positions, fluxes, and qualities of fit. There is a clump of measurements with $r > 0.985$, and a trail of measurements with lower-quality fits. For each of the eight E6 observations, we choose the measurements that have $r > 0.985$ and determine a robust average (x,y) position, flux, and RMS about the average. These measurements of course are not independent, but their agreement gives us a sense of the PSF-subtraction errors.

Figure S9 shows the 21 independent observations for the position in each of the eight E6 exposures. Each $x$ and $y$ observation is plotted on the vertical axis (in blue and red, respectively) and the quality of fit ($r$) is shown on the horizontal axis. The E6 observations correspond to exposures number 19 through 26. The observation with $r > 0.985$ all lie to the right of the dashed vertical black line. We provide the average $x$ and $y$ positions in blue and red respectively, along with the RMS of the various measurements about the mean. The magnitude is given in black, along with the RMS about that mean. These x and y positions are used in our subsequent analysis. Although the many measurements for each exposure from the differently subtracted WD are not statistically independent, the eight measurements *are* independent, since they come from different exposures. Of course, there is a dither, so the raw measurement position must be turned into a position relative the neighboring stars. We use these measurements now to check if the measurements could be affected by the presence of the WD.

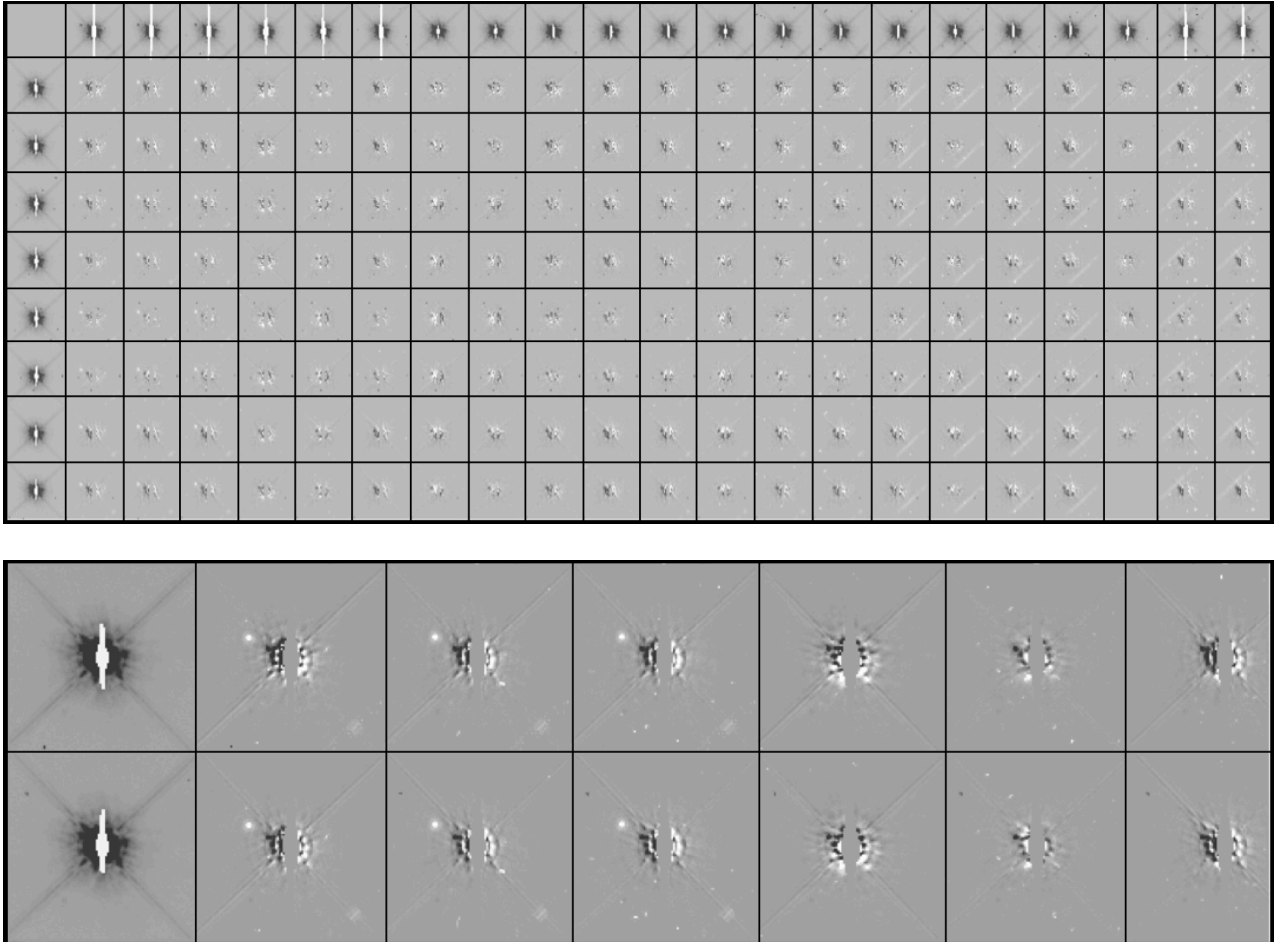

**Fig. S8: Results of different PSF subtractions in E6 observations.** The upper panel shows all the PSF-subtracted raster images. The lower panel shows the top two E6 observations, and the first six subtractions on an expanded scale. The size of each raster image is 25×25 pixels.

Fig. S10 shows how well our method of optimal PSF subtraction works. The left panel shows the image of Stein 2051 A, the second panel shows the WD+target, and the right panel shows the source after the subtraction of the PSF. The contribution of the residual WD PSF at the location of the source is now negligible. The residual fluxes are mostly confined to the saturated pixels around Stein 2051B.

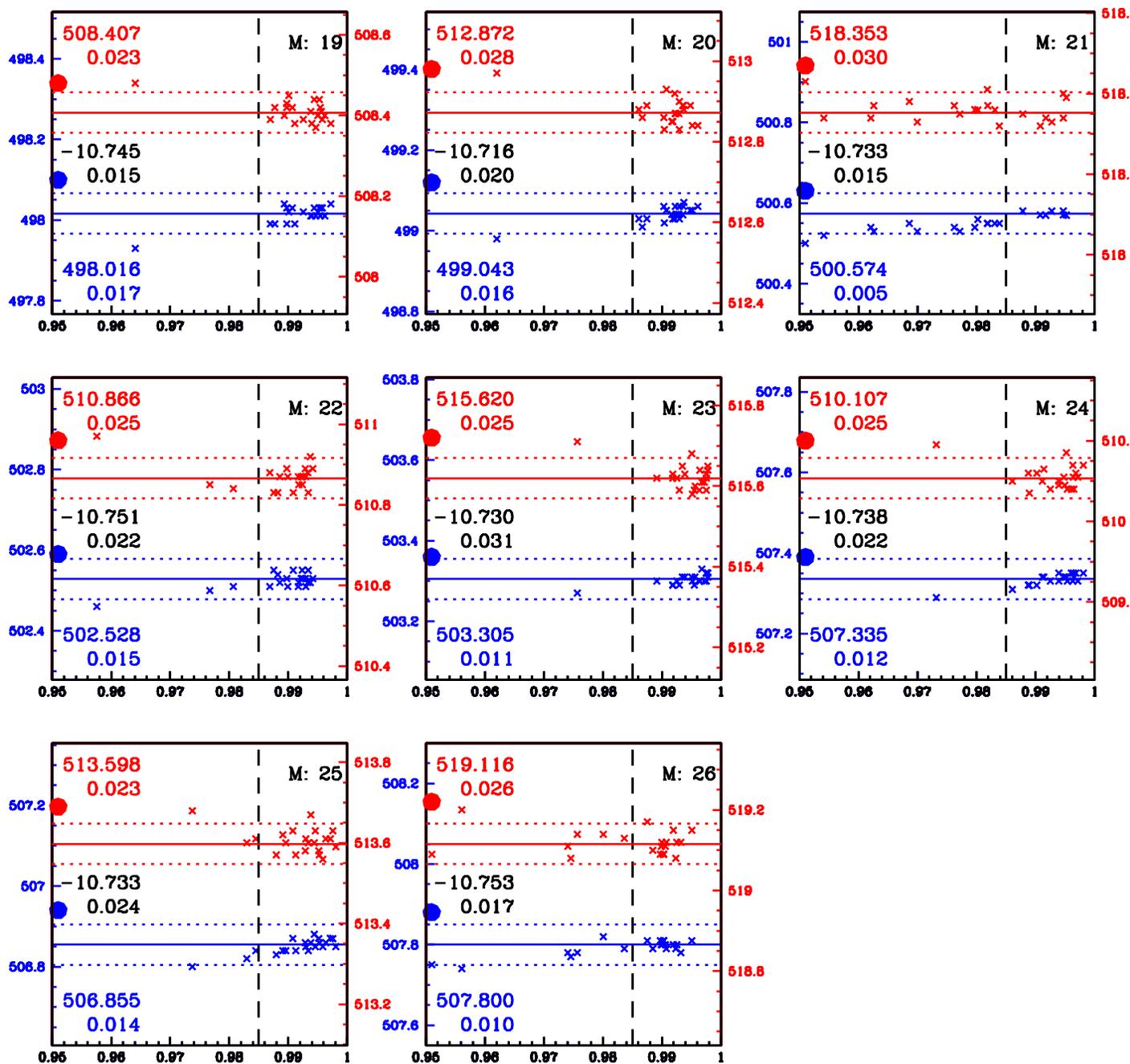

**Fig S9: Source positions with different PSF subtractions.** The vertical axis shows the measured *x* and *y* positions (in blue and red, respectively) and the horizontal axis shows the quality of fit (*r*).

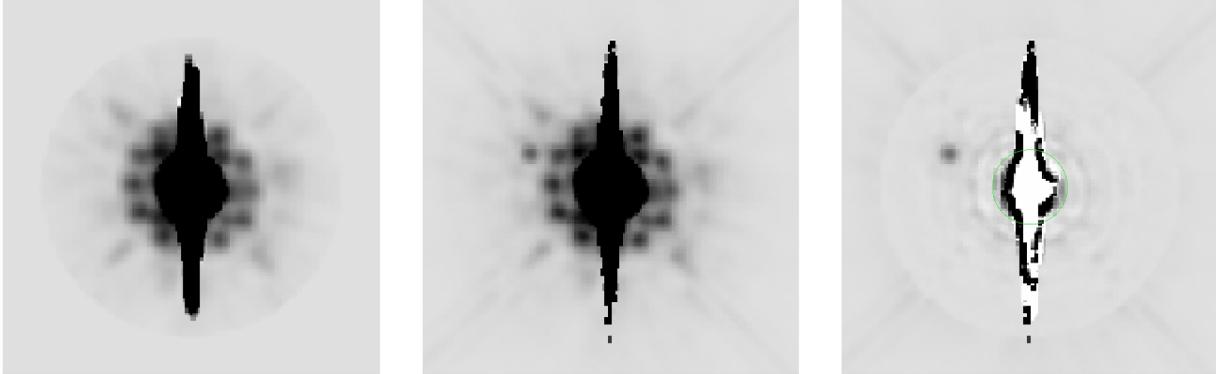

**Fig. S10: The effect of PSF subtraction.** The image on the left shows Stein 2051 A, the M dwarf, the image in the middle shows Stein 2051 B + the source, and the image on the right shows Stein 2051 B + the source after optimal PSF subtraction.

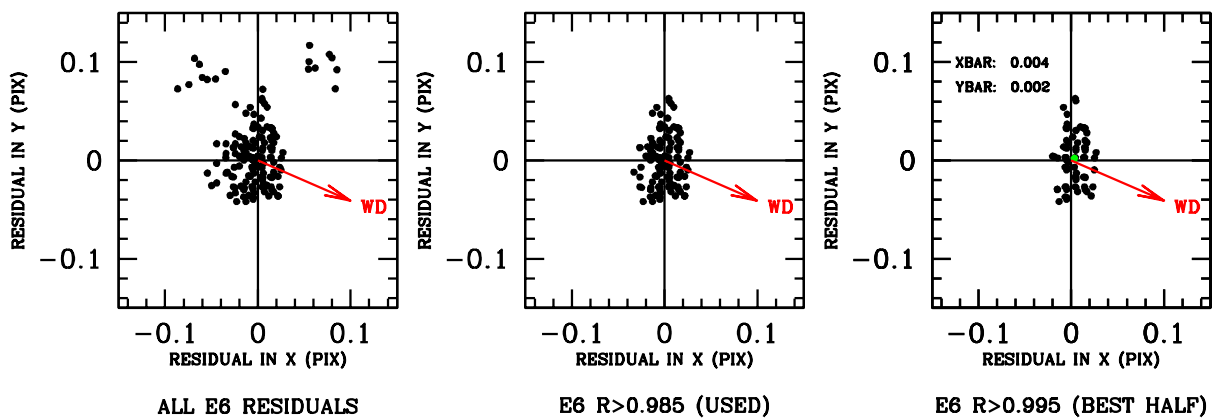

**Fig. S11: Residuals in position measurements relative to the average.** All the points are shown on the left panel, which include some observations affected by cosmic rays. The points we actually used are shown in the middle panel, and the best half of all the points are shown in the right panel.

Figure S11 shows the residuals relative to the averages in the 8 panels in the previous plot. The left panel shows all the points, the middle panel shows the ones we used, and the right panel shows the best half. The position of the WD with respect to the source is shown by a red arrow. These residuals should be an indication of the impact of the WD subtraction. The points clearly don't show any particular stretching towards the WD. If we consider the residuals with the best fits (r>0.995), the offset is <0.005 pixels in the direction of the WD, which is well within our measurement uncertainties.

The fact that we don't see any stretching of the residuals in the direction of the WD is not surprising in retrospect, since the PSF at the distance of the source is more dominated by features than by the radial gradient. However, it was important to confirm that the presence of the WD does not stretch the PSF of the source in the direction towards the WD. The strong radial of the PSF gradient stops at about 2 pixels away from center of the source. Since we use a 3x3 pixel window to fit the target, the gradient does not play much of a role.

**Supplementary Text**

Astrophysical parameters of Stein 2051 B

We determined the distance to the Stein 2051 system by calculating a weighted mean of published parallax measurements of Stein 2051 A and B, including the value determined in the present work (Table S1). The weighted mean is 181.0±0.5 mas, corresponding to a distance of 52±0.01 pc.

The effective temperature of Stein 2051 B (WD 0426+588) had been determined previously (*22*) using a technique of fitting model-atmosphere fluxes to observed broad-band *BVRI* and *JHK* photometry. The *BVRI* and *JHK* magnitudes are converted into monochromatic absolute fluxes, which are then compared with model fluxes, averaged over the same filter bandpasses, using a chi-square minimization method. The two fitting parameters are the effective temperature and the solid angle $\pi(R/D)^2$, where *R/D* is the ratio of the radius of the star to its distance from Earth. The model fluxes are available online (*57*). The assumption was made of a pure helium photospheric composition.

This procedure yielded $T_{eff}$ = 7178±182 K (*22*). However, a minor error was subsequently discovered in the calculations of the cool, pure helium model grid, which has now been updated. With this correction, we now find $T_{eff}$= 7122±181 K. Using the above parallax, the stellar radius is found to be 0.0114±0.0004 $R_\odot$. The corresponding luminosity is log ($L/L_\odot$) = -3.52±0.02.

**Table S1: Absolute parallax measurements [milliarcsec] for Stein 2015 system**

| Stein 2051 A | Stein 2051 B | Source |
|---|---|---|
| 177.9±1.6 | 179.7±1.9 | (*36*) |
| 179.27±3.23 | ... | (*37*) |
| 181.50±0.62 | ... | (*38*) |
| ... | 181.5±1.0 | Present work |

**Movie M1**

Motion of Stein 2051 B.

**Movie M2**

Relativistic Deflection Animation.